\documentclass[11pt]{article}
\usepackage{geometry}                
\usepackage{graphicx}
\usepackage{subfigure}
\usepackage{amssymb, amsmath}
\usepackage{epstopdf}
\newcommand{\BE}{\begin{equation}}
\newcommand{\EE}{\end{equation}}

\title{Analysis of series expansions for non-algebraic singularities}
\author{Anthony J Guttmann}
\date{}                                           

\begin{document}
\maketitle
\begin{abstract}
Existing methods of series analysis are largely designed to analyse the structure of algebraic singularities. Functions with such singularities have their $n^{th}$ coefficient behaving asymptotically as $A \cdot \mu^n \cdot n^g.$ Recently, a number of problems in statistical mechanics and combinatorics have  been encountered in which the coefficients behave asymptotically as $B \cdot \mu^n \cdot \mu_1^{n^\sigma} \cdot n^g,$ where typically $\sigma = \frac{1}{2}$ or $\frac{1}{3}.$  Identifying this behaviour, and then extracting estimates for the critical parameters $B, \,\, \mu, \,\, \mu_1, \,\, \sigma, \,\, {\rm and} \,\, g$ presents a significant numerical challenge. We describe methods developed to meet this challenge.
\end{abstract}

\section{Introduction}
The method of series analysis  has, for many years, been a powerful tool in the study of many problems in statistical mechanics, combinatorics, fluid mechanics and computer science. In essence, the problem is the following: Given the first $N$ coefficients of the series expansion of some function, (where $N$ is typically as low as 5 or 6, or as high as 100,000 or more), determine the asymptotic form of the coefficients, subject to some underlying assumption about the asymptotic form, or, equivalently, the nature of the singularity of the function.

Typical examples include the susceptibility of the Ising model, and the generating function of self-avoiding walks (SAWs). These are believed to behave as 
\begin{equation}\label{generic}
F(z) = \sum_n c_n z^n \sim C \cdot (1 - z/z_c)^{-\gamma}.
\end{equation}
  In the Ising case, for regular two-dimensional lattices, the values of both $z_c$ and $\gamma=7/4$ are exactly known, and the amplitude $C$ is known to more than 100 decimal places. In the SAW case, the value of $z_c$ is only known for the hexagonal lattice \cite{DC10}, and the value of $\gamma=43/32$ is universally believed, but not proved. 

The method of series analysis is used when one or more of the critical parameters is not known. For example, for the three-dimensional versions of the above problems, none of the quantities $C,$ $z_c$ or $\gamma$ are exactly known.
 From the binomial theorem it follows that 
\begin{equation}\label{asymp1}
c_n \sim \frac{C }{\Gamma(\gamma)}\cdot  z_c^{-n} \cdot  n^{\gamma-1}
\end{equation}
 Here $C, \,\, z_c, \,\, {\rm and} \,\, \gamma$ are referred to as the critical amplitude, the critical point and the critical exponent respectively.

The aim of series analysis is to obtain, as accurately as possible, estimates of the critical parameters from the first $N$ coefficients. Since obtaining these coefficients is typically a problem of exponential complexity, the usual consequence is that fewer than 100 terms are known (and in some cases far fewer)\footnote{In the case of the susceptibility of the two dimensional Ising model, polynomial time algorithms for enumerating the coefficients have been developed \cite{O00,C10}, and in that case we have hundreds of terms. Unfortunately, this is a rare situation.}.

There are literally thousands of such problems in statistical mechanics, combinatorics, computer science and fluid mechanics (and other areas) where such a situation arises. 

The methods to extract estimates of the critical parameters from the known series expansion largely fall into two classes. One class is based on the {\em  Ratio method}, initially developed by Domb and Sykes \cite{DS56}, and subsequently refined and expanded by many authors. 

The second is based on analysing a differential equation the solution of which  has an algebraic singularity (\ref{generic}). It is constructed so that the first $N$ terms of the power series expansion of its solution precisely agree with the known expansion coefficients of the underlying problem. The first development of this nature was due to Baker \cite{GAB61}, based on taking Pad\'e approximants of the logarithmic derivative of known series. This was then substantially extended by Guttmann and Joyce \cite{GJ72} who developed the {\em method of differential approximants}, which is still the most successful method in use today for analysing series with algebraic singularities, typified by (\ref{generic}).

While, as noted, many problems have such an algebraic singularity structure, an increasing number of situations have been encountered in which a more complex singularity structure prevails. Those cases are characterised by coefficients with dominant asymptotics of the form
\begin{equation}\label{asymp2}
b_n \sim C \cdot z_c^{-n}\cdot \mu_1^{n^\sigma}\cdot n^{g}.
\end{equation}
That is to say, there is a sub-dominant term $\mu_1^{n^\sigma},$ giving rise to two additional parameters, $\mu_1$ and $\sigma.$ If $\mu_1 > 1,$ we can write down a generic generating function whose coefficients have this asymptotic behaviour, but if $\mu_1 < 1,$ a generic generating function does not appear to be known, (at least not by the author).

An important {\em caveat} to this work is that, as we argue below, the exponent $\sigma$ appearing in  eqn. (\ref{asymp2}) is a simple rational fraction. Indeed in all the situations we've encountered it takes one of only three values, $1/2,$ $1/3$ or $2/3.$ For the Interacting Partially Directed Self-Avoiding Walk (IPDSAW) model, defined in Section \ref{IPDSAW}, and more general models of the collapse transition in interacting walk models \cite{VR00}, there are powerful physical arguments based on the presence of a surface free-energy that can be shown \cite{D93} to give rise to such a term with exponent $\sigma=(d-1)/d,$ where $d$ is the dimensionality of the system. In other cases, such as Dyck paths counted by both length and height, as discussed in Section \ref{sec:ex2}, probabilistic arguments\footnote{R Pemantle, private communication.}, based on the expected behaviour in the scaling limit of the objects being counted, can be used to prove that $\sigma = 1/3.$

The existence of such asymptotic behaviour has been proved in some cases. For example the coefficients of the exponential generating function (EGF) of fragmented partitions \cite{FS09}, and the ordinary generating function (OGF) of Dyck paths, subjected to a compressing force applied to the top-most vertex (discussed below). In other cases one has field theoretical arguments, such as those used by Duplantier and colleagues \cite{DS87,DD88,D93} in their discussion of two-dimensional collapsed dense polymers, and multiple Manhattan lattice walks, and careful numerical work based on series expansions of an exact solution \cite{BGW92} of interacting partially directed self-avoiding walks (IPDSAW) by Owczarek et al. \cite{OPB93}.

Another important observation is that for functions whose coefficients have the asymptotic form (\ref{asymp1}), the underlying generating function has, almost invariably, an algebraic singularity of the form (\ref{generic}). For functions whose coefficients have the asymptotic form (\ref{asymp2}), the underlying generating function can be a well-behaved D-finite function (as in the case of fragmented permutations), or a function with a natural boundary (as in the case of integer partitions), as well as perhaps something in between of which we don't have an example. So while for algebraic singularities one can perhaps carelessly fail to distinguish between the singularity and its asymptotic form, one must be much more careful when discussing series whose coefficients behave like (\ref{asymp2}). For want of a better name, we'll refer to these as non-algebraic singularities, while accepting that that describes a much wider class of singularities than those considered here.
 In this work we develop methods to identify the asymptotic form of the coefficients, assuming it is (\ref{asymp1}) or (\ref{asymp2}). We will have nothing to say about the underlying singularity. 

Very recently, my colleagues and I have come across several situations in which this generic asymptotic behaviour seems to arise. In combinatorics the notoriously unsolved problem of $1324$ pattern-avoiding permutations has, conjecturally, this asymptotic behaviour \cite{CG14}. A number of two-dimensional self-avoiding walk (SAW) problems in which the walk is subject to a compressive force also, conjecturally, have coefficients with this asymptotic form. These include self-avoiding bridges, SAWs  and polygons. In these models we consider the situation in which the bridge/walk/polygon originates in a horizontal line. The two-variable generating function is $$G(x,y) = \sum_{n,h} g_{n,h} x^n z^h,$$ where $g_{n,h}$ is the number of objects of length $n$ with maximal height ($y$-coordinate) $h.$ If $z < 1,$ then squat, broad objects are favoured over tall, slim objects. This then models a compressive force applied to the object. For these models, in the compressed regime, there are physical arguments (as alluded to above) that make plausible the existence of the asymptotic structure (\ref{asymp2}), and these models will be discussed in future publications, currently in preparation\footnote{A number of papers by various subsets of N~ Beaton, A~J~Guttmann, I~Jensen, E~J Janse van Rensburg and S~G Whittington are currently being written.}.

Given the increasingly frequent occurrence of problems where coefficients have such an asymptotic form, it has become pressing to develop numerical techniques to estimate the various critical parameters. That is the purpose of this article. We first outline the two principal methods used to analyse algebraic singularities. We then discuss how these methods behave when applied to the class of non-algebraic singularity we are considering here. Naturally, they fail in this case, but the nature of their failure gives information about the true nature of the singularity.

We then show how the ratio method can be modified and extended to be useful in analysing these non-physical singularities, and how the method of differential approximants can also be applied to provide useful information. In the next section we describe the traditional ratio method. In the following two sections we describe the method of Pad\'e approximants, and then the method of differential approximants (DAs), showing just how precise estimates can be obtained in favourable circumstances. Our main purpose here is to show just how good the DA method is in  estimating the critical parameters of algebraic singularities. By contrast, it performs very poorly when given a series possessing the type of non-algebraic singularity considered here. It is precisely this poor performance that indicates the presence of this type of singularity\footnote{Indeed, this observation was the catalyst for this work. Our series analysis in a problem of compressed polygons we were studying behaved so uncharacteristically badly , we were driven to find out why, and this work is the result.}.

 In Section \ref{sec:nonalg} we discuss this particular non-algebraic asymptotic behaviour, and give a generic OGF that has coefficients with the appropriate asymptotic form (when $\mu_1 \ge 1$). We show how to modify and extend the ratio method so that it can be used in such situations. We then discuss the application of the method of differential approximants to such non-physical singularities, and show how the coefficients can be transformed to new coefficients which behave, to leading order, like those of an algebraic singularity.

In subsequent sections we study three examples, of increasing difficulty, which are known to have non-algebraic singularities with the asymptotic behaviour (\ref{asymp2}) considered here. Our first example is a slightly modified version of the generating function for fragmented permutations. The second is an analysis of Dyck paths subjected to a compressing force at their top vertex, and our final example is that of IPDSAWs. 

We then  conclude by giving a method, or more precisely a number of methods, which collectively provide an effective recipe for analysing series expansions with coefficients of this non-algebraic asymptotic form. Furthermore, the methods provide effective tools for {\em predicting} that the asymptotic form is of the presumed type.
\section{Ratio Method}
The ratio method \index{ratio method} was perhaps the earliest systematic
method of series analysis employed,
and is still a useful starting point, prior to the application of more sophisticated
methods. It was first used by M~F~Sykes in his 1951 D Phil studies, under the supervision
of C~Domb. From equation (\ref{asymp1}),
it follows that the {\it ratio} of successive terms
\begin{equation} \label{ratios}
r_n = \frac{c_n}{c_{n-1}}=\frac{1}{z_c}\left (1 + \frac{\gamma -1}{n} + {\rm o}(\frac{1}{n})\right ).
\end{equation}
 From this idea, it is then natural to plot the successive ratios $r_n$ against $1/n.$
If the correction terms ${\rm o}(\frac{1}{n})$ can be ignored\footnote{For a purely algebraic singularity (\ref{generic}), with no confluent terms, the correction term will be ${\rm O}(\frac{1}{n^2}).$}, such a plot will be linear,
with gradient $\frac{\gamma-1}{z_c},$ and intercept $1/z_c$ at $1/n = 0.$

\begin{figure}
\centerline {
\includegraphics[width=10cm]{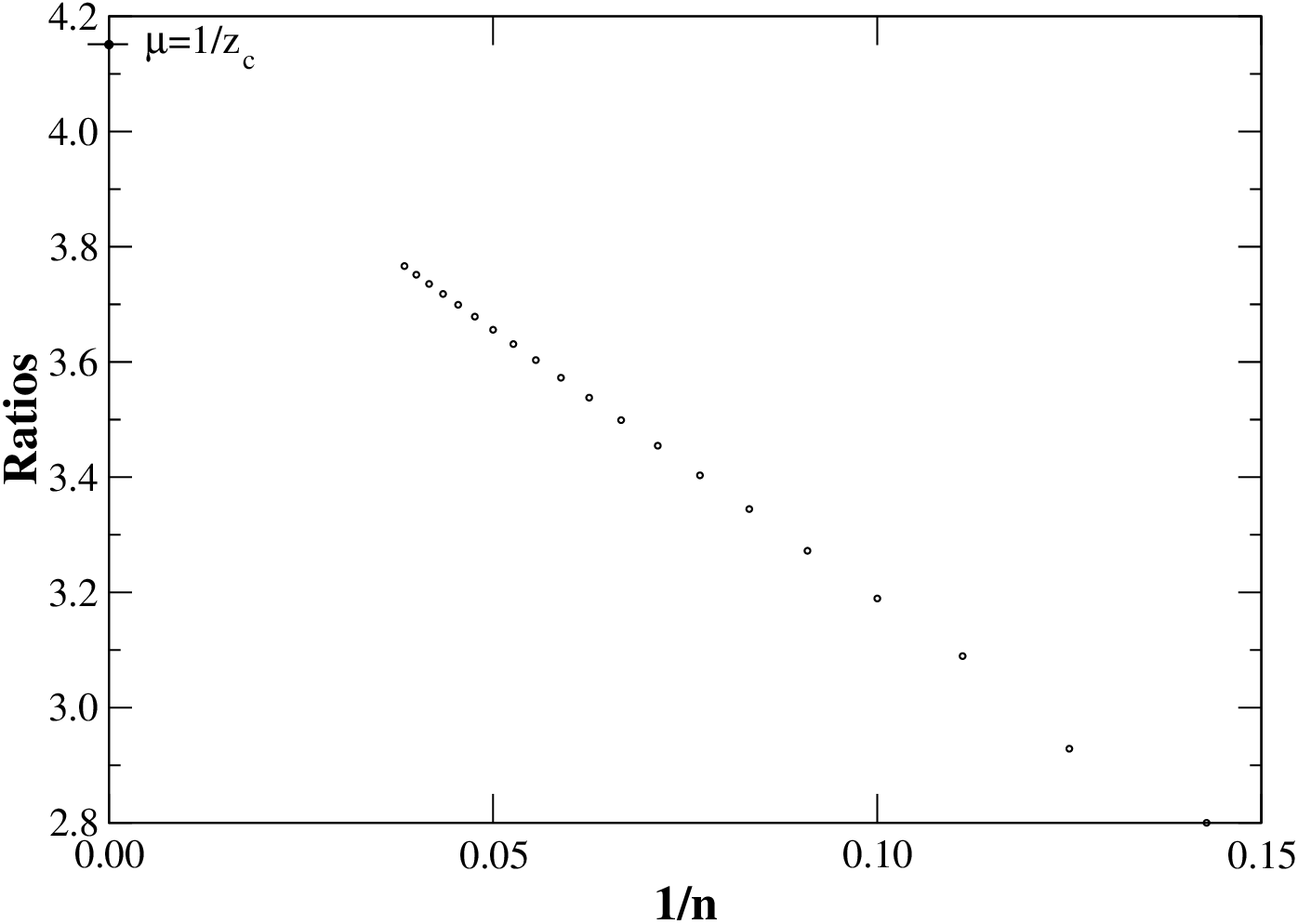}
}
\caption{\label{fig:ratt}
Plot of ratios against $1/n$ for triangular lattice polygons. A straight line through the
last few data points intercepts the Ratios axis at $1/z_c.$
}
\end{figure}

As an example, we apply the ratio method to the generating function of self-avoiding polygons (SAPs) on the triangular lattice. The first few terms in
the generating function, from $p_3$ to $p_{26}$ are:
 2,
    3,
   6,
    15,
   42,
    123,
    380,
   1212,    3966,   13265,
   45144,
   155955,
   545690,
   1930635,
   6897210,
   24852576,
   90237582,
  329896569,
   1213528736,
   4489041219,
  16690581534,
  62346895571,
   233893503330,
   880918093866.
Plotting successive ratios against $1/n$ results in the plot shown in Figure \ref{fig:ratt}.
The critical point is estimated \cite{Jensen04} to be at $z_c \approx 0.240917574\ldots = 1/4.15079722\ldots .$
   \vspace{-0.2in}
\vspace{.2in}
From the figure one sees that the locus of points, after some initial (low $n$) curvature,
becomes linear to the naked eye for $n > 15$ or so, (corresponding to $1/n < 0.067$).
Visual extrapolation to $1/z_c$ is quite obvious. A straight line drawn through the
last $4-6$ data points intercepts the horizontal axis around $1/n \approx 0.13.$ Thus
the gradient is approximately $\frac{4.1508-2.8}{-0.13} \approx -10.39,$
from which we conclude that the exponent $\gamma - 1 \approx -10.39 \cdot z_c \approx -2.50.$ It is believed \cite{N82}
that the exact value is $\gamma = -3/2,$ which is in complete agreement with this
simple graphical analysis.

Various refinements of the method can be readily derived. If the critical point
is known exactly, it follows from equation (\ref{ratios}) that estimators of the exponent
$\gamma$ are given by
$$ \gamma_n=n(z_c\cdot r_n-1)+1 = \gamma +  {\rm o}(1).$$
If $z_c$ is unknown, estimates of the exponent $\gamma$ can be obtained by defining estimators $\gamma_n$ of $\gamma$ and extrapolating these against $1/n.$ Here
\BE \label{eq:exp}
\gamma_n=1+n^2\left ( 1-\frac{r_n}{r_{n-1}} \right )= \gamma +  {\rm o}(1).
\EE
Similarly, if the exponent $\gamma$ is known, estimators of the critical point $z_c$
are given by  $$z_c^{(n)} =  \frac{n+\gamma-1}{n r_n}= z_c + {\rm o}(1).$$
One problem with the ratio method is that if the singularity closest to the origin
is not the singularity of interest (the so-called {\it physical singularity}),
\index{physical singularity} then
the ratio method will not give information about the physical singularity. Worse still,
if the closest singularity to the origin is a conjugate pair of singularities, the ratios will vary dramatically in both
sign and magnitude. To overcome this difficulty G A Baker Jr \cite{GAB61} proposed
the use of Pad\'e approximants applied to the logarithmic derivative of the series expansion.

We should also mention that there exists a vast literature of extrapolation techniques in numerical analysis, and many such methods can be advantageously applied to extrapolate the sequence of ratios in order to estimate the radius of convergence, which is the critical point. Some of these methods, applied to series analysis problems, are discussed in the review \cite{G89}. In particular, the Bulirsch-Stoer algorithm \cite{BS} has been found to be quite powerful, as it allows for the more general situation when convergence is not linear in $1/n.$ (Recall that, for an isolated algebraic singularity, convergence is always linear in $1/n,$ so in that case the Bulirsch-Stoer method affords no advantage). We have used the Bulirsch-Stoer method to estimate the radius of convergence in all the examples with non-algebraic singularities that we consider below.


\section{Pad\'e approximants} 
\label{ana:pade}

The basic idea of using Pad\'e approximants for series analysis is very simple. Given a function $F(z)$
with a simple pole at some point $z_c$ we use the series expansion of $F(z)$
to form a rational approximation to $F(z),$ 

\BE \label{eq:ana_pade}
F(z) = \frac{P_i(z)}{Q_j(z)}
\EE
where $P_i(z)$ and $Q_j(z)$ are polynomials of degree $i$ and $j$ respectively,
whose coefficients are chosen such that the first $i+j+1$ terms in the series expansion
of $F(z)$ are identical to those of the expansion of $P_i (z)/Q_j(z),$ with $Q_j(0)=1$ for uniqueness. Constructing the polynomials only involves solving a system of linear equations.

In order to use the Pad\'e approximation scheme to reliably approximate an algebraic singularity rather than just
a meromorphic functions, 
we must first transform the series into a suitable form. This brings us to the
classic method called Dlog-Pad\'e \index{Dlog-Pad\'e} approximation \cite{GAB61}.
If we have a function with  expected behaviour typical of algebraic singular
points, as given
by equation (\ref{generic}),
then taking the derivative of the logarithm of $F(z)$ gives
\BE \label{eq:ana_Dlog}
\widehat{F}(z)=\frac{\rm d}{{\rm d}z} \log F(z) \simeq \frac{\gamma}{z_c-z} +{\rm O}(1).
\EE
This form is perfectly suited for Pad\'e analysis, as taking the logarithmic derivative has turned the function into a meromorphic function (at least to leading order). We see that an
estimate of the critical point $z_c$ can be obtained from the
roots of the denominator polynomial $Q_j(z)$, while an estimate of the
critical exponent $\gamma$ is
obtainable from the residue of the Pad\'e approximant to $\widehat{F}(z)$ at $z_c$,
that is
\BE \label{eq:ana_resexp}
\gamma \approx \lim_{z\to z_c} (z_c-z)\frac{P_i(z)}{Q_j(z)}.
\EE

Since $\widehat{F}(z) = F'(z)/F(z),$ we see that
forming a Dlog-Pad\'e approximant is simply equivalent to seeking an approximation
to $F(z)$ by solving the first order homogeneous differential equation

$$F'(z)Q_j(z)-F(z)P_i(z) =0.$$
This observation leads us directly to the more powerful and more general method
of {\em differential approximants} by noting that we
can approximate $F(z)$ by a solution to a higher order ODE (possibly inhomogeneous).
 This method was first proposed and developed by
Guttmann and Joyce \cite{GJ72} in 1972, and was subsequently extended to the
inhomogeneous case by Au-Yang and Fisher \cite{FA79} and Hunter and Baker \cite{HB79} in 1979.
The advantage of a higher order ODE is that confluent singularities can be accommodated, as well as a more complicated singularity structure in general. Functions that satisfy such an ODE are called {\em D-finite} or {\em holonomic}.

\section{Differential approximants}
\label{ana:da}

The generating
functions  of many lattice models
in statistical mechanics and combinatorics are often algebraic, or otherwise given by the solution of simple linear ODEs.
This observation (originally made in the context
of the 2-d Ising model) is the origin of the method of {\em differential approximants}.
The basic idea is to approximate a generating function $F(x)$ by solutions
of differential equations with polynomial coefficients. The singular behaviour
of such ODEs is a well known classical mathematics problem
(see e.g. \cite{Forsyth02,Ince27}) and the singular points and
exponents are easily calculated. Even if {\em globally} the function is not describable by a solution
of a such a linear ODE (as is usually the case) one hopes that
{\em locally,} in the
vicinity of the (physical) critical points, the generating
function is still well-approximated by a solution to a linear ODE.

An $M^{th}$-order differential approximant (DA) to a function $F(z)$  is formed by matching
the coefficients in the polynomials $Q_k(z)$ and $P(z)$ of degree $N_k$ and $L$, respectively,
so that the formal solution of the inhomogeneous differential equation
\BE \label{eq:ana_DA}
\sum_{k=0}^M Q_{k}(z)(z\frac{{\rm d}}{{\rm d}z})^k \tilde{F}(z) = P(z)
\EE
agrees with the first $N=L+\sum_k (N_k+1)$ series coefficients of $F(z)$. Constructing such ODEs only involves
solving systems of linear equations. The function
$\tilde{F}(z)$ thus agrees with the power series expansion of the (generally unknown)
function $F(z)$ up to the first $N$ series expansion coefficients.
We normalise the DA by setting $Q_M(0)=1,$ thus leaving us with $N$ rather
than $N+1$ unknown coefficients to find, in order to specify the ODE. The choice of the differential operator $z\frac{{\rm d}}{{\rm d}z}$ in (\ref{eq:ana_DA}) forces the origin to be a regular singular point. The reason for this choice is that most lattice models with holonomic solutions, for example, the free-energy of the two-dimensional Ising model, possess this property.

From the theory of ODEs, the singularities of $\tilde{F}(z)$ are approximated by zeros
$z_i, \,\, i=1, \ldots , N_M$ of $Q_M(z),$ and the
associated critical exponents $\gamma_i$ are estimated from the indicial equation. If there is only a single root at $z_i$  this is just
\BE \label{eq:ana_indeq1}
\gamma_i=M-1-\frac{Q_{M-1}(z_i)}{z_iQ_M ' (z_i)}.
\EE

Details as to which approximants should be used and how the estimates from many approximants are combined to give a single estimate are given in \cite{GJ09}. In the next sub-section  we give an example of the application of the method. 

\subsection{The honeycomb SAP generating function}
\label{ana:hcsap}
In this sub-section we apply the method of differential approximants to the
generating function for self-avoiding polygons (SAPs) on the honeycomb lattice. The generating function $$P(x) = \sum_{n \ge 1} p_nx^{2n}$$ is expected to have a dominant singularity $const. \cdot (1-x^2/x_c^2)^{2-\alpha}.$ On this lattice the critical point is known rigorously \cite{DC10}, and the
 critical exponent and
some universal amplitude ratios are believed to be known exactly. In Table~\ref{tab:ana_HCDA} we have listed the estimates
for the critical point $x_c^2$ and exponent $2-\alpha$ obtained from second- and
third-order DAs. We note that all the estimates are in  agreement
 in that within `error-bars'  they take the same value.
From this we arrive at the  estimate $x_c^2=0.2928932186 \pm 5 \times 10^{-10}$ and $2-\alpha=1.5000004 \pm 1 \times 10^{-6}$.
The final estimates are in perfect agreement with the 
exact values \cite{DC10} $x_c^2=1/\mu^2=1/(2+\sqrt{2})= 0.292893218813\ldots$ and $2-\alpha=3/2$.

\begin{table}
\caption{\label{tab:ana_HCDA}
Critical point and exponent estimates for self-avoiding polygons. Numbers in parentheses give the uncertainty in the last quoted digits.}
\begin{center}
\begin{tabular}{lllll} \hline \hline
 $L$   &  \multicolumn{2}{c}{Second order DA} &
       \multicolumn{2}{c}{Third order DA} \\
\hline
    &  \multicolumn{1}{c}{$x_c^2$} & \multicolumn{1}{c}{$2-\alpha$} &
      \multicolumn{1}{c}{$x_c^2$} & \multicolumn{1}{c}{$2-\alpha$} \\
\hline
 0 & 0.29289321854(19)& 1.50000065(41) &  0.29289321865(12)& 1.50000040(28) \\
 5 & 0.29289321875(21)& 1.50000010(59) &  0.29289321852(48)& 1.50000041(99) \\
10 & 0.29289321855(23)& 1.50000060(48) &  0.29289321878(32)& 1.49999999(97) \\
15 & 0.29289321859(19)& 1.50000054(43) &  0.29289321861(37)& 1.50000035(67) \\
20 & 0.29289321866(15)& 1.50000038(33) &  0.29289321860(21)& 1.50000049(43) \\
\hline \hline
\end{tabular}
\end{center}
\end{table}

\begin{figure}
\begin{center}
\includegraphics[width=11cm]{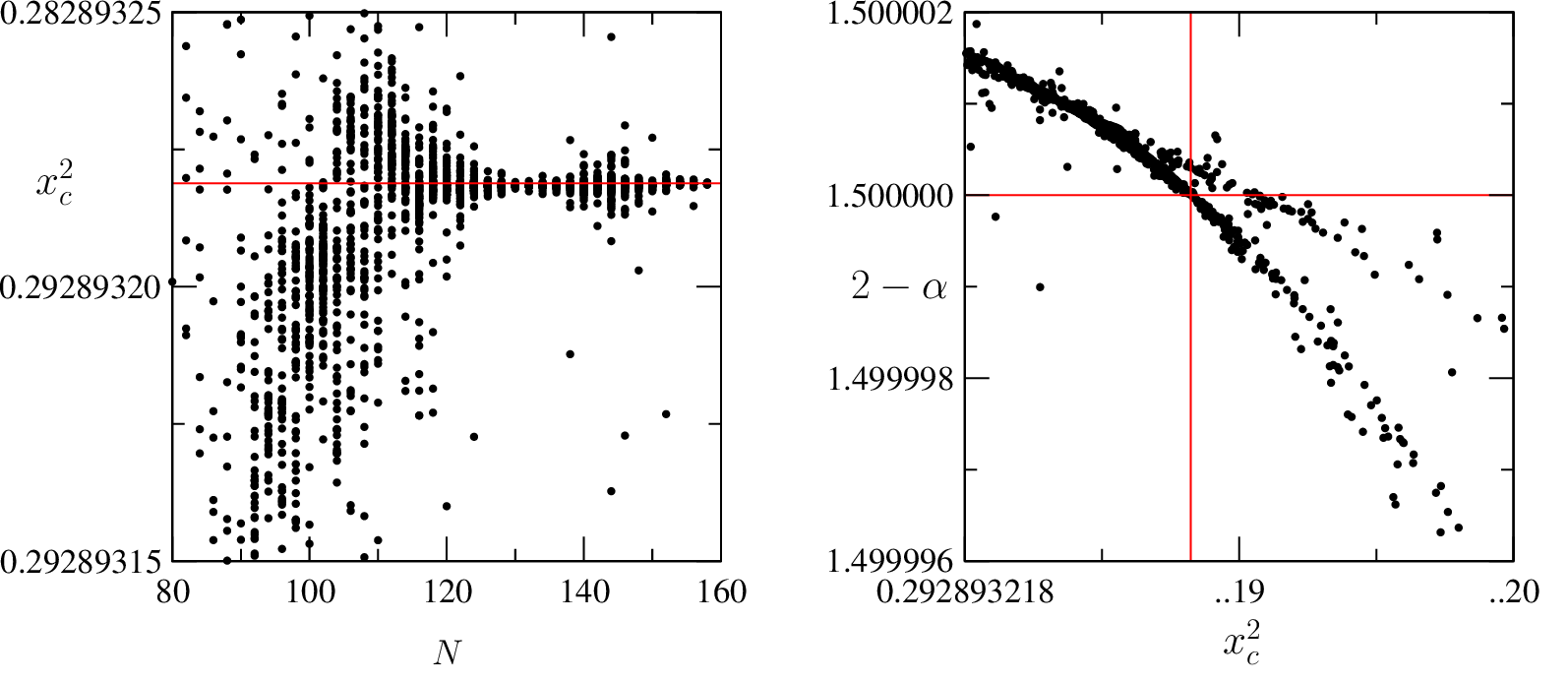}
\end{center}
\caption{\label{fig:ana_HCDA}
Plot of estimates from third order differential approximants
for $x_c^2$ vs. the highest order term used, and the right panel shows $2-\alpha$ vs. $x_c^2$. The straight lines
are the exact predictions.
}
\end{figure}

Not surprisingly, the estimates improve as the number of available series terms increases. This can be seen in the left panel
 of Figure~\ref{fig:ana_HCDA} where  the
estimates  from third-order DAs for $x_c^2$ vs. the highest order coefficient index $N<N_{max}$ used by the DA are plotted.
Each dot in the figure is an estimate obtained from a specific approximant. As can be
seen, the estimates clearly settle down to the conjectured exact value (solid line) as
$N$ is increased, and there is no evidence of any systematic drift at large $N$.

In the right-hand panel we show the variation in the exponent estimates with the critical point
estimates.  Thus if one knows or conjectures either the exponent or critical point, a more precise estimate of the other can be obtained. The `curve' traced out by the estimates passes through the intersection
of the lines given by the exact values. The apparent branching into two arcs is probably spurious.

One of the reasons for giving this example is to show just how successful and precise the method is under favourable circumstances. We argue that, by 
contrast, when the method behaves badly, with poorly converged estimates of the radius of convergence and wildly varying exponent estimates, that
this is a signal that the underlying singularity is not an algebraic singularity. In the next section we discuss other types of singularities that give rise to a more complicated asymptotic form.

\section{Functions with non-algebraic singularities.}\label{sec:nonalg}

A number of solved, and, we claim, unsolved problems that arise in lattice critical phenomena and algebraic combinatorics have coefficients with a more complex asymptotic form, with a sub-dominant  term ${\rm O}(\mu_1^{n^\sigma})$ rather then ${\rm O}(n^g).$ In fact the sub-sub dominant term is of ${\rm O}(n^g)$. Perhaps the best-known example of this sort of behaviour is the number of partitions of the integers -- though in that case the leading exponential growth term $\mu^n$ is absent (or equivalently $\mu=1$).

There are a number of models in mathematical physics that also have a more complex asymptotic structure, of the type we are discussing here. In particular, Duplantier and Saleur \cite{DS87} and Duplantier and David \cite{DD88} studied the case of {\em dense} polymers in two dimensions, and found the partition functions had the asymptotic form
$$Q_n \sim const. \cdot \mu^n \cdot \mu_1^{n^\sigma} \cdot n^{g}.$$ In \cite{OPB93}, Owczarek, Prellberg and Brak investigated an exactly solvable model of interacting partially-directed self-avoiding walks (IPDSAW), for which the solution had previously been given by Brak, Guttmann and Whittington in \cite{BGW92}. In particular they analysed a 6000 term series expansion for IPDSAWs in the collapse regime, and estimated $\sigma = 1/2,$ $g = -3/4,$ while $\mu_1$ was found to at least 6 digit accuracy. From \cite{BGW92} the value of $\mu$ is exactly known. Subsequently Duplantier \cite{D93} pointed out that $\sigma = 1/2$ is to be expected, not only for IPDSAWs, but also for SAWs in the collapsed regime. In subsection \ref{IPDSAW} we show that the methods we develop below can give good results using only about 100 terms (occasionally 200), rather than 6000 used in \cite{OPB93}. This is of practical importance, as for many unsolved problems one typically only has 20-200 terms available.

An example from combinatorics is given by the exponential generating function (EGF) of fragmented permutations\footnote{A fragmented permutation is an unordered collection of non-empty sub-permutations of a given permutation. For example, there are three fragmented permutations of two elements: $\{1,2\},$ $\{2,1\}$ and $\{1\},\{2\}.$}  \cite{FS09} which is $$F(z) =\exp\left ( \frac{z}{1-z} \right ).$$ Then with $F_n=[z^n]F(z),$ we have \cite{FS09}, p563 $$F_n \sim \frac{e^{2\sqrt{n}}}{2\sqrt{\pi e}\cdot n^{3/4}}.$$

This follows from Wright \cite{W32, W49} who calculated the leading asymptotic form of the expansion of
\BE \label{eq:W}
F(z) = (1-\mu z)^{-\beta}\exp \left ( \frac{A}{(1-\mu z)^\rho} \right ), \,\,\, A>0, \,\, \rho >0.
\EE
For $\rho \le 1,$ Wright's saddle-point analysis yields
\BE
[z^n]F(z) \sim \mu^n \frac{N^{\beta-1-\rho/2}\exp(A(\rho+1)N^\rho)}{\sqrt{2\pi A\rho(\rho+1)}},
\EE
 with $N := \left ( \frac{n}{A\rho} \right )^{\frac{1}{\rho+1}}.$

This asymptotic form can be written as 
\BE \label{can1}
 B \cdot \mu^n \cdot \mu_1^{n^\sigma} \cdot n^g,
\EE
 where  $ \mu_1=\exp(A^{\frac{\rho}{\rho+1}}\cdot \rho^{-\frac{1}{\rho+1}}),$ so in particular $\mu_1 > 1.$ Also $\sigma = \frac{\rho}{\rho+1},$ and $g=\frac{\beta - 1 - \rho/2}{\rho+1}.$

For this situation, with $\mu_1 > 1,$ the term involving $\mu_1^{n^\sigma}$ rapidly dominates the term $n^g,$ for any value of $g.$ However if $\mu_1 < 1,$ the term $\mu_1^{n^\sigma}$ is eventually smaller than the contribution of the term $n^g.$ For the situation $\mu_1 <1,$ we are unaware of any analogue of Wright's expansion. That is to say, we do not know what generic closed form expression, analogous to (\ref{eq:W}), has an asymptotic expansion of the form (\ref{can1}) with $\mu_1 < 1.$


In the remainder of this paper we develop numerical methods to analyse functions whose coefficients have the asymptotic form given in eqn. (\ref{can1}), based on extensions of the ratio method and the method of differential approximants. We then take three examples of functions of increasing complexity with coefficients that are known to behave asymptotically as in eqn. (\ref{can1}) and see how successful or otherwise the methods are.

\subsection{Ratio method for non-algebraic singularities.}

If the coefficients of some generating function behave as 

\BE \label{eq:an}
b_n \sim B \cdot \mu^n \cdot \mu_1^{n^\sigma} \cdot n^g,
\EE
 then the ratio of successive coefficients $r_n = b_n/b_{n-1},$ is
\begin{multline} \label{eq:rn}
r_n = \mu \left (1 + \frac{\sigma \log \mu_1}{n^{1-\sigma}} + \frac{g}{n} + \frac{\sigma^2 \log^2 \mu_1}{2n^{2-2\sigma}} + \frac {(\sigma-\sigma^2)\log \mu_1+2g\sigma \log \mu_1}{2n^{2-\sigma}} \right . \\
 \left . {}+ \frac{\sigma^3 \log^3 \mu_1}{6n^{3-3\sigma}} +{\rm O}(n^{2\sigma-3}) + {\rm O}(n^{-2}) \right ).
\end{multline}

In the examples considered in this paper, as well as other examples encountered, $\sigma$ takes the simple values $1/2$ or $1/3.$
When $\sigma = \frac{1}{2},$ (\ref{eq:rn}) specialises to
\BE \label{eq:half}
r_n = \mu \left (1 + \frac{ \log \mu_1}{2\sqrt{n}} + \frac{g+\frac{1}{8}\log^2 \mu_1}{n} + \frac{\log^3\mu_1+(6+24g)\log \mu_1 }{48n^{3/2} } + {\rm O}(n^{-2}) \right ),
\EE 
and when $\sigma = \frac{1}{3},$ to
\BE \label{eq:third}
r_n = \mu \left (1 + \frac{ \log \mu_1}{3{n^{2/3}}} + \frac{g}{n} + \frac{\log^2\mu_1}{18n^{4/3}}+ \frac{(2+6g)\log \mu_1 }{18n^{5/3} } + {\rm O}(n^{-2}) \right ).
\EE

So given a series, if one applies the ratio method and finds the ratio plots are not linear, and can be linearized by plotting the ratios against $1/n^{1-\sigma},$ with $\sigma =1/2$ or $1/3,$ then this suggests that the asymptotic form of the coefficients could well be of the type considered here.

From (\ref{eq:rn}), one sees that 
\BE \label{eq:rsigma}
(r_n/\mu-1) \sim const. n^{\sigma-1}.
\EE
 Accordingly, a log-log plot of $\log(r_n/\mu-1)$ versus $\log{n}$ should be linear, with gradient $\sigma-1.$ We would expect an estimate of $\sigma$ close to that which linearised the ratio plot.

Estimating $\sigma$ this way requires knowledge of, or at worst a very precise estimate of, the growth constant $\mu.$ While $\mu$ is exactly known in the three examples considered below, more generally $\mu$ is not known, and must be estimated, along with all the other critical parameters. In order to estimate $\sigma$ without knowing $\mu,$ we can use one (or both) of the following estimators:

From eqn. (\ref{eq:rn}), it follows that 
\BE \label{eq:sig1}
r_{\sigma_n} = \frac{r_n}{r_{n-1}} \sim 1 + \frac{(\sigma-1)\log{\mu_1}}{n^{2-\sigma}} + {\rm O}(1/n^2),
\EE
so $\sigma$ can be estimated from a log-log plot of $\log(r_{\sigma_n}-1)$ against $\log{n}.$

Another estimator of $\sigma$ follows from eqn. (\ref{eq:an}),
\BE \label{eq:sig2}
a_{\sigma_n} = \frac{b_n^{1/n}}{b_{n-1}^{1/(n-1)}} \sim 1 + \frac{(\sigma-1)\log{\mu_1}}{n^{2-\sigma}} + {\rm O}(1/n^2),
\EE
so again $\sigma$ can be estimated from a log-log plot of $\log(a_{\sigma_n}-1)$ against $\log{n}.$ 

While these two estimators are equal to leading order, they differ in their higher-order terms. And indeed, as shown below, which of the two is more informative varies from problem to problem.

\subsection{Direct fitting for non-algebraic singularities} \label{direct}
Another, perhaps obvious, idea is to try and fit the critical parameters directly to the assumed asymptotic form. The assumed asymptotic form is $$b_n \sim B\cdot \mu^n \cdot \mu_1^{n^\sigma} \cdot n^g$$ Therefore

\BE \label{logcan1}
\log {b_n} \sim \log{B} + n \log{\mu} + n^\sigma \log{\mu_1} + g \log {n}.
\EE
So if $\sigma$ is known, or assumed, we have four unknowns in this linear equation. It is then straightforward to solve the linear system
$$\log {b_k} =  c_1k  + c_2 k^\sigma  +c_3 \log {k}+c_4$$ for $k=n-2,\, n-1, \, n, \, n+1$ with $n$ ranging from $3$ to $N-1,$ where $N$ is the power of the highest known series coefficient. Then $c_1$ estimates $\log(\mu),$ $c_2$ estimates $\log(\mu_1)$, $c_3$ estimates $g$ and $c_4$ gives estimators of $\log{B}.$ An obvious variation arises in those cases where, say, $\mu$ is known. Then one can solve  
$$b_k - k \log{\mu} =  c_2 k^\sigma  +c_3 \log {k} +c_4$$ 
from three successive coefficients, as before increasing the order of the lowest used coefficient by one until one runs out of coefficients.

\subsection{Using the method of differential approximants}
In this sub-section we investigate the use of the method of differential approximants in the analysis of series with asymptotic coefficients of the form (\ref{asymp2}). We will see in our first example -- a modified version of the generating function for fragmented permutations -- that the EGF is in fact holonomic, satisfying a first-order linear ODE. So an appropriately chosen differential approximant will solve this problem completely, based on only a few terms in the series expansion. So this is not a  testing example.

Our second example, that of height-weighted Dyck paths, is more typical. If one simply applies the method of differential approximants, the results, discussed in Section \ref{sec:ex2}, suggest that the generating function is not well-approximated by a linear ODE of the assumed type -- and hence that the singularity is not likely to be algebraic. This behaviour is typical of those cases where the singularity is not of the assumed algebraic type. That is to say, in such cases one typically sees imprecise and inaccurate estimates of the critical point, unrealistic values of the associated critical exponent, and sometimes a concentration of other critical points along the real axis. This behaviour is characteristic of the situation in which the differential approximants are trying unsuccessfully to represent the singularity(ies) of the coefficients of the underlying generating function.

Earlier in this section we discussed Wright's function (\ref{eq:W}), which generates coefficients of the asymptotic form (\ref{asymp2}) considered here. However, as discussed, Wright's function only generates asymptotic forms for its coefficients when $\mu_1 > 1.$ For those situations when $\mu_1 < 1,$  Wright's function does not generate coefficients of the required asymptotic form. Indeed, if one asks the natural question,``what OGF has coefficients with asymptotic behaviour $const. \cdot 4^n \cdot \mu_1^{n^{1/3}} \cdot n^g$ where $\mu_1 < 1?$'' the answer seems to be unknown\footnote{It is certainly not given by Wright's OGF with $A < 0,$ for in that case the coefficients actually change sign with a known periodicity.}.

We can (partially) side-step this difficulty by constructing an OGF with coefficients  which are just the reciprocals of the original coefficients. For if $c_n \sim const. \cdot \lambda^n \cdot \mu_1^{n^{\sigma}} \cdot n^g,$
then $d_n = 1/c_n \sim const. \cdot \lambda^{-n} \cdot (1/\mu_1)^{n^{\sigma}} \cdot n^{-g}.$ So if $\mu_1 < 1$ the coefficients are now of a form given by the asymptotic expansion of  Wright's function. Unfortunately, this new generating function maps singularities that were previously beyond the radius of convergence in the original series closer to the origin than the mapped physical singularity, and if there are several of these, they dominate the asymptotic behaviour.

Note that if we take the logarithmic derivative of Wright's function (\ref{eq:W}), we obtain
\BE \label{eq:ldW}
{\tilde F}(z)=\frac{d}{dz} \log{F(z)} = \frac{F'(z)}{F(z)} = \frac{\beta \mu}{1-\mu z} + \frac{A \rho \mu}{(1-\mu z)^{\rho+1}}.
\EE
${\tilde F}(z)$ now has algebraic singularities, and so might be amenable to analysis by the method of differential approximants. That is to say, we might expect the logarithmic derivative of the OGF of the reciprocal series to behave as a function with algebraic singularities at $z = 1/\mu,$ and with exponents $-1$ (a simple pole) and a dominant branch point with exponent $-(\rho+1).$ Numerical experiments show that these transformations -- taking the logarithmic derivative of the OGF with reciprocal coefficients -- substantially improve the performance of the differential approximants method for the analysis of non-physical singularities of the assumed type, but, while useful, are not as accurate as we need for a reliable method. Fortunately, we have developed a different transformation that is more effective.

\subsection{Transforming series to remove the factor $\mu_1^{n^{\sigma}}.$}\label{sec:transform}

As noted above, the method of differential approximants is of limited use in analysing  series which are not dominated by an algebraic singularity.
For those series with coefficients with the asymptotic form considered here, it is the presence of the $\mu_1^{n^\sigma}$ term that is responsible for the lack of applicability of the method. However we can manipulate the series to remove the offending term, and then use this powerful method.  From eqn. (\ref{logcan1}) one has, when $\sigma = 1/2,$

$$\log {b_n} = \log{B} + n \log{\mu} + \sqrt{n} \log{\mu_1} + g \log {n} + {\rm O}(1/\sqrt{n}).$$ Then with $\tilde{b}_n = \log{b_n}/\sqrt{n},$ we can form new coefficients $c_n:$

\BE  \label{eq:renorm1}
c_n=2n^{3/2}(\tilde{b}_n - \tilde{b}_{n-1}) = (2g-\log B) + n\log(\mu)-g\log(n) + {\rm O}(1/\sqrt{n}).
\EE
Exponentiating these coefficients, we have 
$$d_n = \exp(c_n) = D\cdot \mu^n \cdot n^{-g}\cdot (1 + {\rm O}(1/\sqrt{n}),$$ where $D=e^{2g}/B.$

When $\sigma = 1/3,$ one has

$$\log {b_n} = \log{B} + n \log{\mu} + n^{1/3} \log{\mu_1} + g \log {n} + {\rm O}(1/n^{1/3}).$$ Then defining $\tilde{b}_n =\log{ b_n}/n^{1/3},$ one has:

\BE  \label{eq:renorm2}
c_n=\frac{3}{2}n^{4/3}(\tilde{b}_n - \tilde{b}_{n-1}) = \frac{3g-\log B}{2} + n\log(\mu)-\frac{g}{2}\log(n) + {\rm O}(1/n^{1/3}).
\EE
So in this case
$$d_n = \exp(c_n) = D\cdot \mu^n \cdot n^{-g/2}\cdot (1 + {\rm O}(1/n^{1/3}),$$ where $D=e^{3g/2}/\sqrt{B}.$

In this way we have transformed the series to one whose coefficients, $d_n$ behave asymptotically, at least to leading order, like a function with an algebraic singularity. We can therefore analyze the series with transformed  coefficients $d_n$ by the method of differential approximants (DAs). Note however that the correction terms are O$(1/n^\sigma),$ whereas for an isolated algebraic singularity they are O$(1/n)$, so one can't expect the standard methods, like the method of differential approximants, to perform as well with the transformed series as, say, the example in Section \ref{ana:hcsap}.

We can also apply other standard techniques to the analysis of the transformed series. The ratios of successive terms ($d_n$) of the transformed series when plotted against $1/n$ are now linear, but as the simple ratio method doesn't give us a particularly accurate estimate of $\mu,$ we don't give the results here. Rather, we extrapolate the ratios of the coefficients of the transformed series using the Bulirsch-Stoer algorithm, with parameter $w=1,$ as appropriate for an expected correction term O$(1/n).$ 

In order to estimate the critical exponent $g,$ we also tried the simple ratio method, extrapolating estimators $g_n = n^2\left ( 1-\frac{r_n}{r_{n-1}} \right )$ of the exponent against $1/n,$ as described at eqn. (\ref{eq:exp}). 

In summary, it is clearly useful to transform the original series as described by eqns. (\ref{eq:renorm1},\ref{eq:renorm2}) and apply the standard methods of series analysis.

In the next three sections we will consider three problems whose coefficients have the assumed asymptotic form, $$b_n \sim B \cdot \mu^n \cdot \mu_1^{n^\sigma} \cdot n^g.$$ In all cases we first try to estimate the value of $\sigma$ and $\mu.$ After determining the value of $\sigma,$ the estimate of $\mu$ is refined. Next we estimate the other critical parameters $\mu_1$ and $g.$ Finally the amplitude term $B$ is estimated.

\section{Example 1. Modified fragmented permutations.}\label{sec:ex1}


We take as our first example a minor variant of the EGF of fragmented permutations and consider
\BE \label{eq:ex1}
F_1(z) =\exp\left ( \frac{2z}{1-2z} \right ),
\EE
Then with $f_n=[z^n]F_1(z),$ 
\BE \label{asympfp}
f_n\sim 2^{n-1}\frac{e^{2\sqrt{n}}}{\sqrt{\pi e}\cdot n^{3/4}}.
\EE
 $F_1(z)$ is clearly holonomic, satisfying the simple ODE $$(1-2z)^2F_1'(z)=2F_1(z), \,\,\, F_1(0)=1,$$ but we don't make use of this in the subsequent analysis.

\begin{figure}[t!]
\centering
\subfigure[Plot of ratios of coefficients of (\ref{eq:ex1}) against $\frac{1}{n}$.]{\includegraphics[width=7.3cm]{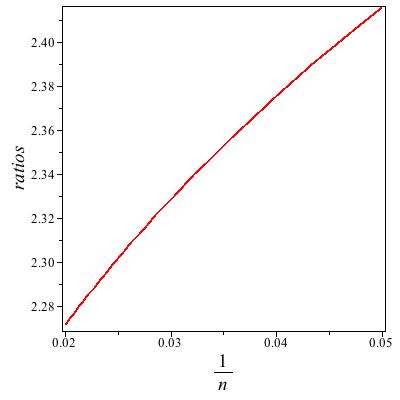}} 
\subfigure[Plot of ratios of coefficients of (\ref{eq:ex1}) against $\frac{1}{\sqrt{n}}$.]{\includegraphics[width=7.3cm]{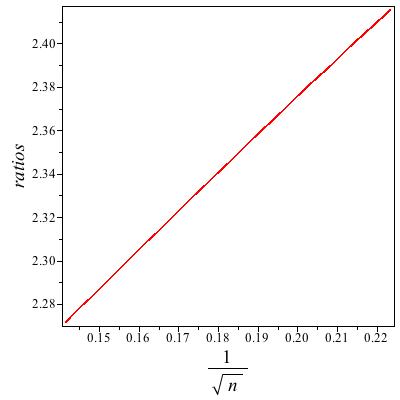}} 
\caption{}
\label{fig:rat1}
\end{figure}

We have generated the series expansion of (\ref{eq:ex1}) up to the coefficient of $z^{50}$ to attempt an analysis\footnote{Of course it is trivial to generate vastly longer series, but a series of 50 terms is not atypical in those frequent cases where the coefficients have to be calculated by some algorithm of exponential complexity.}.
Applying the ratio method (\ref{ratios}) to these coefficients, the resulting plot is shown in Figure \ref{fig:rat1}(a). That is to say, we plot the ratios $r_n = \frac{f_n}{f_{n-1}}$ against $\frac{1}{n}.$  Unlike the case of an algebraic singularity, with ratio plots shown in Figure \ref{fig:ratt}, here one sees considerable curvature in the  plot. This is the hallmark of the type of non-algebraic singularity we are considering here.

We show in Figure \ref{fig:rat1}(b) the same ratios, but now plotted against $\frac{1}{\sqrt{n}}.$ 
This plot appears to be linear, and also to be approaching the expected limit of 2 as $n \to \infty.$ Plotting the ratios against $\frac{1}{n^{2/3}}$ (not shown) also gives a plot that looks almost as linear as Figure \ref{fig:rat1}(b), so trying to distinguish the correct value of $\sigma$ in this way is not very precise. The best we can do is to estimate that $\sigma$ is in the range $[0.4,0.7].$

\begin{figure}[t!]
\centering
\subfigure[Log-log plot  of $(r_n/\mu-1)$ against ${n}$.]{\includegraphics[width=7.3cm]{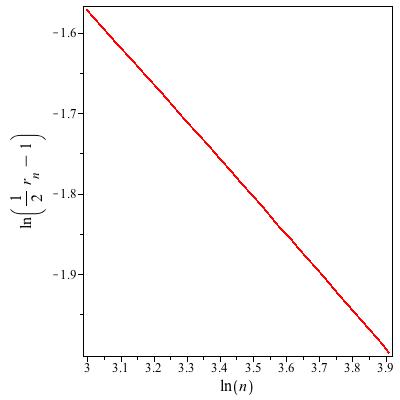}} 
\subfigure[Estimators of $\sigma$ from gradient ratios against ${1/n}$.]{\includegraphics[width=7.3cm]{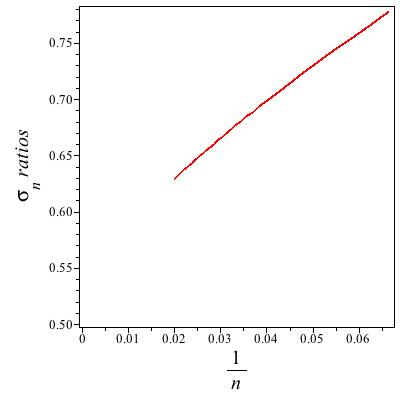}} 
\caption{}
\label{fig:sig1}
\end{figure}

In order to more accurately estimate the value of $\sigma,$  we show in Figure \ref{fig:sig1}(a) a log-log plot of $(r_n/\mu-1)$ versus $\log{n}.$ This is seen to be linear, and the gradient, calculated from the last two ratios, $r_{49}$ and $r_{50}$ is -0.475. Recall that this gradient should be $\sigma-1.$ If one accepts that $\sigma$ is a simple rational number, the value of $1/2$ is inescapable.

If we didn't know the value of $\mu,$ we could estimate $\sigma$ from the gradient of log-log plots of $r_{\sigma_n},$ see eqn. (\ref{eq:sig1}), or $a_{\sigma_n},$ see eqn. (\ref{eq:sig2}). It turns out that they are equally good, and we show in Figure \ref{fig:sig1}(b) the estimate of $\sigma$ given by the gradient of the line joining the points $r_{\sigma_k}$ and $r_{\sigma_{k-1}},$ as $k$ ranges from 15 to 50. There is some curvature in the plot, but clearly a limit of $0.5$ is attainable. We extended the plot to 250 terms (not shown), and the curvature increased, making the known limit 0.5 totally evident. 

The point we want to make here is that if one wants to identify $\sigma$ as a simple fraction, likely to be $1/2$ or $1/3,$ then we have good evidence that it is $1/2.$ This can then be used in subsequent analysis.

Assuming that $\sigma=1/2,$ we next refine the estimate of $\mu.$
We could linearly extrapolate the ratio plot in Figure \ref{fig:rat1}(b) which can be seen to be, plausibly, going to a value around 2 on the ordinate, and we might guess that the value was exactly 2. However, more generally $\mu$ does not take an integral, nor perhaps even an algebraic, value, so it needs to be estimated quite precisely. It is therefore necessary to use an extrapolation algorithm which can accommodate the expected asymptotic behaviour of the ratios. 

The Bulirsch-Stoer algorithm \cite{BS} is such an algorithm, as it  extrapolates sequences that behave as $s_n \sim s_\infty + c/n^w.$ The parameter $w$ is given by the user. In this example, we set $w=1/2,$ and extrapolate the first 50 ratios. The method produces rows of extrapolants that take into account successively higher powers of terms of order $n^{-w}$ as well as terms of order $n^{-m},$ where $m=1,2,\ldots.$ Typically the first few rows behave smoothly, while higher order rows become erratic. We retain only those lower order rows which behave smoothly. In the example given here in Table \ref{tab:BS1}, the first six rows behave smoothly -- by which we mean monotonically. This is unusually good behaviour. Frequently rather fewer rows are monotonic. There is a breakdown of monotonicity in the last row. One would estimate from this table that the limiting value was 1.9999999, and it wouldn't be considered unreasonable to conjecture that the limit is exactly 2.

\begin{table}
\caption{\label{tab:BS1}
Last seven entries in each row of the table of Bulirsch-Stoer extrapolants with $w=1/2.$ Each successive row is the result of a successively higher degree of extrapolation. The available number of coefficients for extrapolation is N (50 in this example). The highest order estimates, and presumably most precise, are all in the last column.}
\begin{center}
\begin{tabular}{llllllll} \hline \hline
 L   &  T(L,N-L-6) & T(L,N-L-5) & T(L,N-L-4)& T(L,N-L-3) & T(L,N-L-2) & T(L,N-L-1) & T(L,N-L) \\
 
\hline

\hline
1 & 2.04480268 &2.04389959& 2.04303343 &2.04220194 &2.04140304 &2.04063483 &2.03989554 \\
2 &1.99676653 &1.99688812 &1.99700239 &1.99710994 &1.99721130 &1.99730695 &1.99739734 \\
3 &2.00011650& 2.00011119 &2.00010622 &2.00010156 &2.00009719 &2.00009309 &2.00008924 \\
4 & 2.00000401& 2.00000347 &2.00000299 &2.00000258 &2.00000222 &2.00000190 &2.00000163 \\
5 & 1.99999280 &1.99999360 &1.99999429 &1.99999488 &1.99999539 &1.99999583 &1.99999622 \\
6 & 1.99999862& 1.99999870& 1.99999878 &1.99999885 &1.99999891 &1.99999897 &1.99999902\\
7 &1.99999991& 1.99999990 &1.99999989 &1.99999989 &1.99999989 &1.99999990 &1.99999990 \\

\hline \hline
\end{tabular}
\end{center}
\end{table}

We will continue the analysis assuming $\sigma = 1/2$ and $\mu=2$ in order to estimate the other parameters, $\mu_1$ and $g$ in the asymptotic form (\ref{can1}). From (\ref{eq:half}), one has 
\BE \label{half}
r_n/2 = 1 + \frac{ \log \mu_1}{2\sqrt{n}} + \frac{g+\frac{1}{8}\log^2 \mu_1}{n} + {\rm O}(n^{-3/2}). 
\EE
In order to estimate $\mu_1$ and $g,$ we solve, sequentially, the pair of equations 
\BE \label{eq:fit1}
r_j/2 = 1 + \frac{ c_1}{\sqrt{j}} + \frac{c_2}{j}, 
\EE
for $ j=k-1$ and $j=k,$ with $k$ ranging from 1 up to 50.

\begin{figure}[t!]
\centering
\subfigure[Estimates of parameter $c_1$ of (\ref{eq:fit1}) against $\frac{1}{n}$.]{\includegraphics[width=7.3cm]{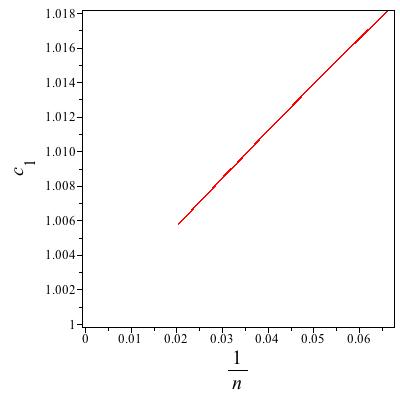}} 
\subfigure[Estimates of parameter $c_2$ of (\ref{eq:fit1}) against $\frac{1}{\sqrt{n}}$.]{\includegraphics[width=7.3cm]{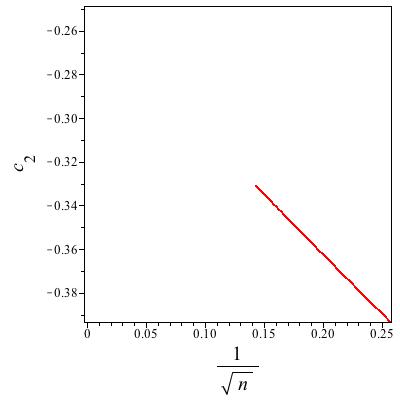}} 
\caption{}
\label{fig:fit1}
\end{figure}

The results are shown in figures \ref{fig:fit1}(a) and \ref{fig:fit1}(b), giving estimates of the parameters $c_1$ and $c_2$ respectively. The first neglected term in (\ref{half}) is O$(n^{-3/2})$ which is O$(1/n)$ smaller than the term with coefficient $c_1,$ so $c_1$ is plotted against $1/n.$ By a similar argument, $c_2$ is plotted against $1/\sqrt{n}.$ A simple extrapolation, literally with a straight-edge, gives the estimates $c_1\approx 1.00$ and $c_2 \approx -0.25.$ From (\ref{half}), $c_1=\log{\mu_1}/2$ and $c_2=g+\log^2{\mu_1}/8.$ Hence we estimate $\log{\mu_1} \approx 2,$ and $g \approx -0.75.$ As it happens, one sees from eqn. (\ref{asympfp}) that these values are exact.

We next tried the idea of direct fitting to the coefficients, as described in Subsection \ref{direct}. Recall that this involves fitting the logarithm of the coefficients to the assumed form and solving successive quartets of equations.
Still using just 50 terms in the generating function (\ref{eq:ex1}), we estimate $c_1\approx0.6932,$ implying $\mu\approx 2.0001,$ (recall that it is exactly 2), $c_2=1.999,$ implying $\mu_1 \approx \exp(1.999),$ (recall that it is exactly $\exp(2)$), $c_3\approx -0.77,$ compared to the exact value $-0.75,$ and $c_4 \approx -1.8$ implying $B \approx 0.16,$ compared to the exact value $B = 0.17109\ldots .$ 

These estimates were obtained quite simply by plotting the successive estimates of each parameter against $1/n$ and visually extrapolating. In each case, without wishing to be too precise, we expect errors to be confined to the last quoted digit.

Fitting to three parameters, imposing the fact that $\mu=2$ is known (or guessing it from the results of the above analysis), the remaining parameters are estimated with significantly improved precision. We estimate $c_2 \approx 1.9995,$ $c_3 \approx -0.75$ and $c_4 \approx -1.77,$ (the exact value is $1.76556\ldots$).

Next we apply the method of differential approximants to the transformed series, as described in Section \ref{sec:transform}. The approximants are found to be well converged, and we estimate $x_c = 1/\mu \approx 0.49999989,$ which differs from the exact value in the 7th decimal place, and $g \approx -0.749.$ These are quite close to the exact values $1/2$ and $-3/4$ respectively. However, this is not a particularly testing example, as the original generating function is holonomic.

We also extrapolated the ratios of the coefficients of the transformed series using the Bulirsch-Stoer algorithm, with parameter $w=1,$ as appropriate for an expected correction term O$(1/n),$ which arises when taking ratios.  We estimate $\mu \approx 2.00000060,$ compared to the exact value $2.0$. 

We can also use the transformed series to directly estimate the exponent $g,$ either imposing prior knowledge of the growth constant $\mu$ or not.  In this instance we did not assume the value of $\mu$ was known, and so extrapolated estimators $g_n = n^2\left ( 1-\frac{r_n}{r_{n-1}} \right )$ of the exponent $g$ against $1/n,$ as described by eqn. (\ref{eq:exp}). In this way we estimated $g \approx -0.755,$ compared to the exact value $-3/4.$

Finally, to estimate the amplitude, we did the most obvious thing and divided the coefficient $f_n$ by the terms we've already identified in the asymptotics form of the coefficients.
That is, we calculated the sequence
$$B_n = \frac{f_n \cdot n^{3/4}}{2^n \cdot \mu_1^{\sqrt{n}}}$$ with $\mu_1=\exp(2).$ Extrapolating the first 50 values $B_n$ against $1/n^{3/4}$ gave a straight line which could be extrapolated, just with a straight-edge, to give the estimate $B \approx 0.1705.$ The exact value is $B=0.171099\ldots $.

So for this rather simple example we see that the suite of methods we have developed combine to give  good numerical estimates of the critical parameters in the asymptotic form of the coefficients. We emphasise that the estimates are predicated on correctly identifying the exponent $\sigma.$

\section{Example 2. Dyck paths enumerated by maximum height.}\label{sec:ex2}
As the second example, we consider the problem of Dyck paths enumerated not just by length, but also by height, which we define to be the maximum vertical distance of a Dyck path from the horizontal axis. Let $d_{n,h}$ be the number of Dyck paths of length $2n$ and height $h,$ so the OGF is $$D(x,y) = \sum_{n,h} d_{n,h} x^{2n}y^h.$$ Then 
\BE
[x^{2n}]D(x,y)=\sum_{h=1}^n d_{n,h}y^h.
\EE
For $y > 1,$ $$D(x,y) \sim \frac{const.}{x_c(y)^2 - x^2},$$ where $x_c(y)=\frac{y}{(y+1)},$ and the constant is $y$-dependent. For $y=1,$ the well-known result is $$D(x,1)=\frac{1 - \sqrt{1-4x^2}}{2},$$ and for $y < 1$ the solution is usually given as an infinite sum of algebraic functions, from which the asymptotic behaviour is difficult to extract. However, it is possible to do so    \cite{BM14}\footnote{I posed this problem at an Oberwolfach meeting in March 2014. Within hours Robin Pemantle confirmed the subdominant behaviour $\mu_1^{n^{1/3}},$ and within 24 hours Brendan McKay gave the complete solution of the dominant asymptotic behaviour given above. Subsequently Nick Beaton derived the sub-dominant term.}, and with constants $A=2^{5/3}\pi^{5/6}/\sqrt{3},$  $C=3 \left (\frac{\pi}{2}\right )^{2/3}$ and $r=-\log{y},$ this is
\BE \label{Dycka}
[x^{2n}]D(x,y)=\frac{(1-y)}{y^2}r^{1/3}A 4^n n^{-5/6} e^{-Cr^{2/3}n^{1/3}}\left (1 +  {\rm O}(n^{-1/3}) \right ).
\EE
So for $y < 1$ we see that coefficients of Dyck paths, indexed by length and height,  behave as (\ref{can1}), with $B=\frac{(1-y)}{y^2}r^{1/3}A$, $\mu=4,$ $\mu_1=\exp(-Cr^{2/3}) $, $\sigma=\frac{1}{3},$ and $g=-\frac{5}{6}.$

In the next subsection we will attempt to determine the critical parameters, assuming the coefficients have the generic asymptotic behaviour (\ref{asymp2}) from the analysis of the series, with $y$ chosen to be 0.5, and using just 50 terms in the series (we actually generated 2500).

\subsection{Numerical analysis of Dyck path series}

Applying the method of differential approximants to the original series, the (very poorly converged) approximants suggest the presence of a singularity at $x_c \approx 0.2511$ (rather than $0.25000$), and with  critical exponent in the range $[6,8],$ which is both an unlikely value and a very imprecise one. Furthermore, the approximants suggest that there are other singularities on the real axis at $x \approx 0.256,$ (with an exponent around 15),  $x \approx 0.286,$ (with an exponent around -12), $x \approx 0.328,$ (with an exponent around -1.7) and poles at $x \approx 0.381,$ $x \approx 0.500,$ $x \approx 0.643,$ and $x \approx 1$\footnote{The exact solution does indeed appear to have singularities along the real axis $x>1/4,$ and which are dense along that ray.}. So this is our first indication that the singularity is non-algebraic. Accordingly, we test for the plausibility that the appropriate asymptotic form of the coefficients are given by eqn. (\ref{can1}).

\begin{figure}[t!]
\centering
\subfigure[Plot of ratios of coefficients of height-weighted Dyck paths with $y=0.5$ against $\frac{1}{n}$.]{\includegraphics[width=7.3cm]{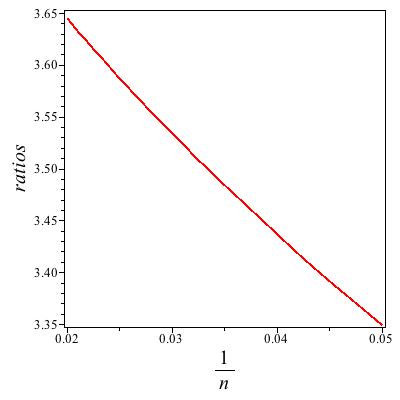}} 
\subfigure[Plot of ratios of coefficients of height-weighted Dyck paths with $y=0.5$ against $\frac{1}{n^{2/3}}$.]{\includegraphics[width=7.3cm]{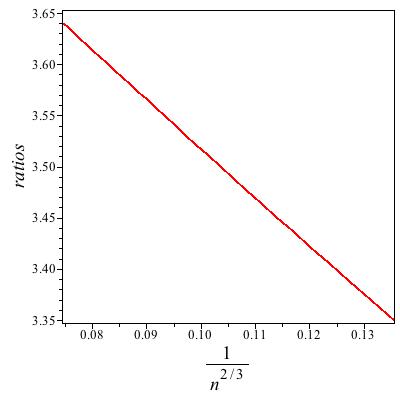}} 
\caption{}
\label{fig:rat2}
\end{figure}

We assume that we don't know the asymptotic form (\ref{Dycka}), but just have the first 50 terms in the expansion, for $y=0.5.$ We repeat the analysis used for example 1 above. We first try a simple ratio plot, the result of which is shown in Figure \ref{fig:rat2}(a).  Some curvature is evident, though not as much as in Figure \ref{fig:rat1}(a), which is not surprising as from (\ref{eq:third}) we expect the ratios to become linear when plotted against $1/n^{2/3}$ whereas in the case of fragmented permutations, the appropriate abscissa was $1/\sqrt{n}.$ In Figure \ref{fig:rat2}(b) we show the ratios plotted against $1/n^{2/3},$ which looks visually linear, and also to be approaching the expected limit of 4. Plotting the ratios against $1/\sqrt{n}$ looks almost as linear, and similarly to the previous example, this crude linearity test only allows us to estimate the value of $\sigma$ to be in the range $[0.4,0.7].$

\begin{figure}[t!]
\centering
\subfigure[Log-log plot  of $1-r_n/4$ against $n$.]{\includegraphics[width=7.3cm]{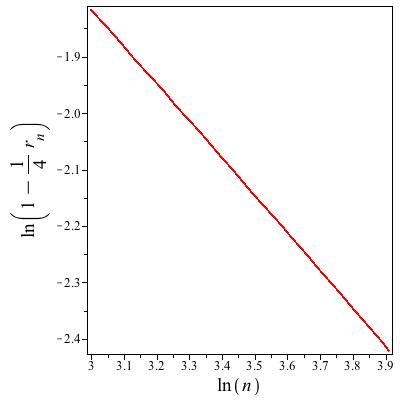}} 
\subfigure[Estimators of $\sigma$ from $a_{\sigma_n}$ ratios against ${1/n}$.]{\includegraphics[width=7.3cm]{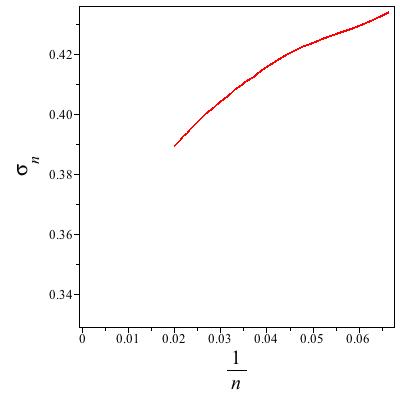}} 
\caption{}
\label{fig:grad2}
\end{figure}

As in the previous example, in order to better estimate the value of $\sigma$ we show in Figure \ref{fig:grad2}(a) a log-log plot of $(1-r_n/\mu)$ against $\log{n}.$ This is seen to be linear, and the gradient, calculated from the last two ratios, $r_{49}$ and $r_{50}$ is -0.675. If one accepts that $\sigma$ is likely to be a simple rational number, the value of $1/3$ is the most compelling guess. (Recall that the gradient of this plot should be $\sigma-1$)\footnote{A more detailed analysis can be conducted, in which the estimates of the gradient formed from increasing successive pairs of ratios, $r_n$ and $r_{n-1}$ are extrapolated against $1/n$, and this does indeed give a value around 0.667, but we don't consider that refinement necessary for this example.}.

Alternatively, if we didn't know the value of $\mu,$ we could estimate $\sigma$ from the gradient of log-log plots of $r_{\sigma_n},$ see eqn. (\ref{eq:sig1}), or $a_{\sigma_n},$ see eqn. (\ref{eq:sig2}). It turns out that estimators from (\ref{eq:sig1}) are decreasing below $1/3,$ only turning around after some 300 terms. However estimators from (\ref{eq:sig2}) are quite informative, and we show in Figure \ref{fig:grad2}(b) the estimate of $\sigma$ given by the gradient of the line joining the points $a_{\sigma_k}$ and $a_{\sigma_{k-1}},$ as $k$ ranges from 15 to 50. There is some curvature in the plot, but clearly a limit of $1/3$ is quite plausible. 

Again we see that if one wants to identify $\sigma$ as a simple fraction, likely to be $1/2$ or $1/3,$ then we have good evidence that it is $1/3.$ This can then be used in subsequent analysis.

In the subsequent analysis we assume that $\sigma=1/3$ in this case. We next require a good estimate of $\mu,$ which from ratio plots we know to be around 4. As in the preceding example, we can extrapolate the ratios using the Bulirsch-Stoer algorithm, this time with parameter $w=2/3.$ The results are given in Table \ref{tab:BS2}. The first four rows behave smoothly -- by which we mean monotonically. The monotonicity breaks down in the fifth row. One would estimate from this table that the limiting value was around $4.001$, and one might conjecture that the limit is exactly 4. If one uses 100 terms instead of 50, the last entries are around 4.0008 and slowly declining. For the remainder of the analysis we assume $\mu = 4.000.$ (Not much changes if we use $4.001$).

\begin{table}
\caption{\label{tab:BS2}
Last seven entries in each row of the table of Bulirsch-Stoer extrapolants. Each successive row is the result of a successively higher degree of extrapolation. The available number of coefficients for extrapolation is $N$ (50 in this example). The highest order estimates, and presumably most precise, are all in the last column, (apart from the last entry).}
\begin{center}
\begin{tabular}{llllllll} \hline \hline
 L   &  T(L,N-L-6) & T(L,N-L-5) & T(L,N-L-4)& T(L,N-L-3) & T(L,N-L-2) & T(L,N-L-1) & T(L,N-L) \\
 
\hline

\hline
1 &4.05166625 &4.05040592& 4.04920422& 4.04805741& 4.04696196& 4.04591457& 4.04491214  \\
2 &4.01339530 &4.01305370 &4.01273785& 4.01244288& 4.01216455& 4.01189944 &4.01164488 \\
3 &4.00367431 &4.00401802 &4.00425109 &4.00437641& 4.00440515& 4.00435326 &4.00423894\\
4 & 4.00762463 &4.00735417 &4.00682191 &4.00600406& 4.00485767& 4.00334352 &4.00147659 \\
5 & 4.00761475 &4.00567628& 4.00495234 &4.00459579 &4.00442333 &4.00438206 &4.00447112  \\
\hline \hline
\end{tabular}
\end{center}
\end{table}

In order to estimate $\mu_1$ and $g,$ recall that from (\ref{eq:third}), it follows that 
\BE \label{twothirds}
r_n/4 = 1 + \frac{ \log \mu_1}{3n^{2/3}} + \frac{g}{n} + {\rm O}(n^{-4/3}). 
\EE

As in the previous example, we solve, sequentially, the pair of equations 
\BE  \label{eq:fit2}
r_j/4 = 1 + \frac{ c_1}{j^{2/3}} + \frac{c_2}{j}, 
\EE
for $ j=k-1$ and $j=k,$ with $k$ ranging from 1 up to 50.

\begin{figure}[t!]
\centering
\subfigure[Estimates of parameter $c_1$ of (\ref{eq:fit1}) against $\frac{1}{n^{2/3}}$.]{\includegraphics[width=7.3cm]{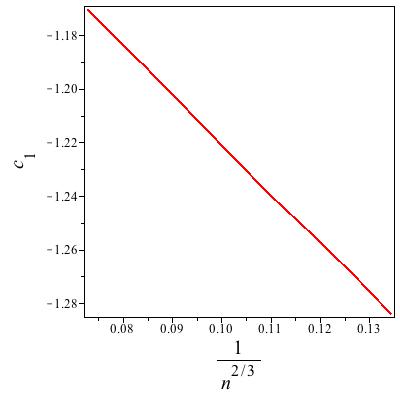}} 
\subfigure[Estimates of parameter $c_2$ of (\ref{eq:fit1}) against $\frac{1}{n^{1/3}}$.]{\includegraphics[width=7.3cm]{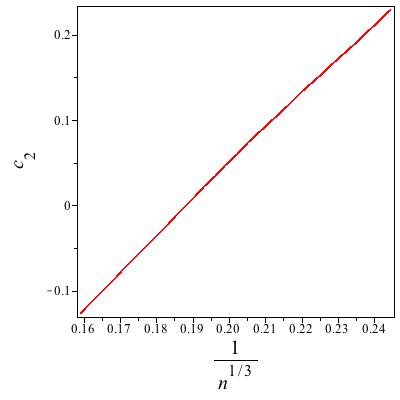}} 
\caption{}
\label{fig:fit2}
\end{figure}

The results are shown in figures \ref{fig:fit2}(a) and \ref{fig:fit2}(b), giving estimates of the parameters $c_1$ and $c_2$ respectively. The first neglected term in equation (\ref{twothirds}) is O$(n^{-4/3})$ which is O$(1/n^{2/3})$ smaller than the term with coefficient $c_1,$ so $c_1$ is plotted against $1/n^{2/3}.$ By a similar argument, $c_2$ is plotted against $1/n^{1/3}.$ A simple extrapolation, literally with a straight-edge, gives the estimate $c_1\approx -1.05.$ The plot for $c_2$ exhibits some curvature, and the best we can estimate is  $c_2 \approx -1.$ From (\ref{eq:third}), $c_1=\log{\mu_1}/3$ and $c_2=g.$ Hence we estimate $\log{\mu_1} \approx -3.15,$ and $g \approx -1.$ The exact values are $\log {\mu_1} = -3.175\ldots$ and $g=-5/6.$ If we take 100 terms in the expansion instead of the 50 that we've used, this method gives the more accurate results $\log {\mu_1} = -3.171\ldots$ and $g=-0.83.$ As we expect critical exponents to be simple rational fractions, the exact value $g=-5/6$ may well be guessed.

Assuming $\sigma = 1/3$ and directly fitting to the remaining parameters, as described in Subsection \ref{direct} above, we estimate
$c_1 \approx 1.3868,$ implying $\mu \approx 4.002$ rather than the exact value of 4, $c_2 \approx -3.28$  rather than the exact value $-3.175\ldots$, $c_3$ is in the range $[-0.9,-0.7],$ compared to the exact value of $-5/6,$ and $c_4 \approx 1.8$ rather than the exact value of $2.1308\ldots$.

If, in addition, we assume that $\mu = 4$ and fit to the remaining three parameters, we find $c_2 \approx -3.20$ rather than the exact value $-3.175\ldots$, $c_3 \approx -0.78,$ compared to the exact value of $-5/6,$ and $c_4 \approx 1.95$ rather than the exact value of $2.1308\ldots$.

We next considered the transformed series (\ref{eq:renorm2}). The differential approximants, applied to the transformed series, while useful, are not as well converged as those in the previous example, with estimates of $x_c$ differing from the exact value in the 5th decimal place, allowing the useful estimate $x_c \approx 0.24998.$ The corresponding exponent estimate is  $g \approx -0.80,$ which can be compared to the correct value $-0.83333\ldots$.

We extrapolated the ratios of the coefficients of the transformed series using the Bulirsch-Stoer algorithm, with parameter $w=1.$  We estimate $\mu \le 4.00036,$ compared to the exact value of $4.0.$ This is more precise than the same analysis applied to the original series.

As in the previous example, we estimated the exponent $g$ by extrapolating estimators $g_n = n^2\left ( 1-\frac{r_n}{r_{n-1}} \right )$  against $1/n,$ as described at eqn. (\ref{eq:exp}).  For this example we estimate $g\approx -0.84,$ compared to the exact value of $-5/6.$ 

Finally, to estimate the amplitude, we did as with the first example and  divided the coefficients by the terms we've identified in the asymptotic form of the coefficients.
That is, we calculated the sequence
$$B_n = \frac{[x^{2n}]D(x,y) \cdot n^{5/6}}{4^n \cdot \mu_1^{n^{1/3}}}$$ with $\mu_1$ taken to be in the range [0.0405,0.043], from the various estimates found above. The exact value is $\mu_1=0.04179\ldots$.
Extrapolating the first 50 values $B_n$ against $1/n^{3/4}$ gave a straight line which could be extrapolated, just with a straight-edge, to give an estimate of $B$ in the rather broad range $[7.3,9.6].$ This large variation is due entirely to the uncertainty in the value of $\mu_1.$ Using the correct value of $\mu_1$ leads to the estimate $B \approx 8.4$. The exact value is $B=8.42208\ldots $.

This example displays behaviour typical of that which we have encountered in other problems, such as SAWs, SAPs and bridges subject to a force. It can be seen that the methods we have developed can clearly identify the nature of the singularity, and also provide good estimates of the various critical parameters, provided a sufficient number of coefficients is known.

\section{Example 3. Interacting partially directed self-avoiding walks}\label{IPDSAW}

For our third and final example, we consider IPDSAW. These are random walks on the square lattice with both west steps and immediate reversals forbidden. The two constraints immediately imply that the paths are self-avoiding. Paths are counted by length, and by the number of monomer-monomer interactions, which occur between adjacent sites that are not consecutive vertices of the walk. The appropriate OGF is $$G(x,y) = \sum_{n,h} c_{n,m} x^{n}y^m,$$ where $c_{n,m}$ is the number of $n$-step IPDSAWs with $m$ monomer-monomer interactions. Then 
\BE
[x^{n}]G(x,y)=\sum_{m=1}^n c_{n,m}y^m.
\EE
This model was solved in \cite{BGW92}. Let $$g_0=1+\sum_{j=1}^\infty \frac{x^{2j}(x-q)^j q^{j(j+1)/2}}{\prod_{i=1}^j (xq^i-x)(xq^i-q)}$$ and 
$$g_1=x+x\sum_{j=1}^\infty \frac{x^{2j}(x-q)^j q^{j(j+1)/2}q^j}{\prod_{i=1}^j (xq^i-x)(xq^i-q)},$$ where $q=xy.$ Then for $y \ne 1,$ with $a=x^2(2-4x)$ and $b=x^2(6-4x),$ the solution is
\BE\label{IPsol}
G(x,y)=\frac{2xg_1-ag_0}{bg_0-2xg_1}.
\EE

The asymptotic form of the coefficients is difficult to extract from (\ref{IPsol}), but based on an analysis of a 6000 term series, Owczarek, Prellberg and Brak \cite{OPB93} conjectured the asymptotic form numerically as $B\cdot\mu^n\cdot\mu_1^{\sqrt{n}} \cdot n^{-3/4},$ where both $\mu > 1$ and $\mu_1 < 1$ depend on the monomer-monomer interaction strength $y,$ in the collapsed regime $y > y_c \approx 3.383$.

Recently, Nguyen and P\'etr\'elis \cite{NP13} have given a more probabilistic exposition of this problem, which has the advantage that the term $\mu_1^{\sqrt{n}}$ in the asymptotic form of the coefficients in the collapsed regime is seen as a natural consequence of the law governing a symmetric random walk. As an aside, we remark that Pemantle's argument (footnote 9) for a term of the form $\mu_1^{n^{1/3}}$ arising in the Dyck path case just discussed is a consequence of the law for reflected Brownian bridges

We expanded (\ref{IPsol}) to obtain 100 terms in the series. It was necessary to obtain somewhat longer series than in our previous examples, as the low order terms involve no monomer-monomer interactions, and it is not until about length 20 that a significant number of interactions occur. As shown in \cite{BGW92}, the tricritical point occurs at $(x_c,yc)=(1/y_c,y_c)$ where $y_c \approx 3.382975\ldots.$ For $y > y_c$ there is a line of critical points lying on the hyperbola $x y = 1.$ As long as we choose a value of $y > y_c,$ we are in the so-called collapsed regime, where the coefficients have the asymptotic form (\ref{can1}). For simplicity we have chosen $y=5,$ so the generating function $G(x,5)$ will have a critical point at $x_c= 1/5,$ so $\mu=5$ in eqn. (\ref{can1}).

In this example, constructing differential approximants to the original series gives very poorly converged results, which as discussed above is  an indication that the underlying OGF does not have a dominant algebraic singularity. The approximants suggest that the critical point is around $0.205$ (rather then $0.2$ exactly), with an exponent in the range $[5,8].$ Again, this large numerical value for the exponent, and its imprecision, suggests that a non-algebraic singularity is dominant.

\begin{figure}[t!]
\centering
\subfigure[Plot of ratios of coefficients of IPDSAWs with $y=5$ against $\frac{1}{n}$..]{\includegraphics[width=7.3cm]{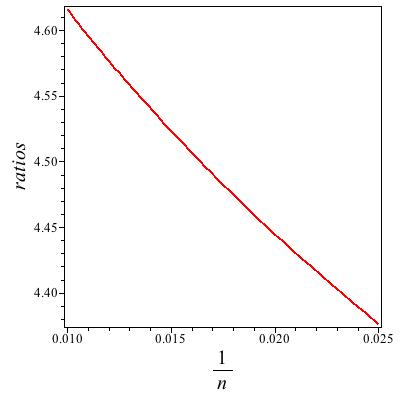}} 
\subfigure[Plot of ratios of coefficients of IPDSAWs with $y=5$ against $\frac{1}{\sqrt{n}}$.]{\includegraphics[width=7.3cm]{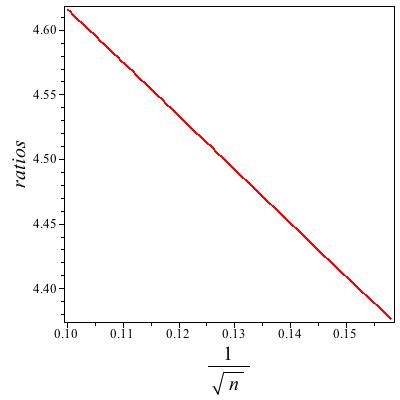}} 
\caption{}
\label{fig:rat3}
\end{figure}

As before, we first plot the ratios of successive terms against $1/n,$ as shown in Figure \ref{fig:rat3}(a).  Some curvature is evident. We next plot the same ratios in Figure \ref{fig:rat3}(b) against $1/\sqrt{n},$ and the plot is seen to be visually linear, implying $\sigma \approx 1/2.$

\begin{figure}[t!]
\centering
\subfigure[Log-log plot  of $(1-r_n/5)$ against $n$.]{\includegraphics[width=7.3cm]{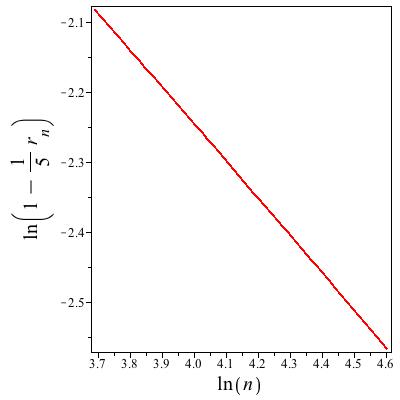}} 
\subfigure[Estimators of $\sigma$ from $r_{\sigma_n}$ ratios against ${1/n}$.]{\includegraphics[width=7.3cm]{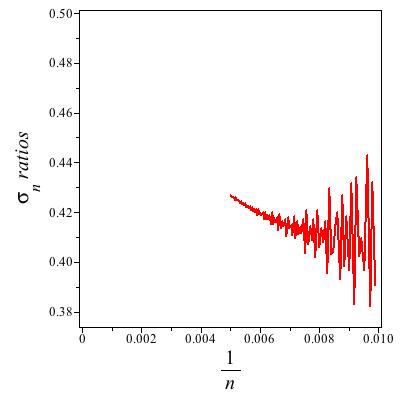}} 
\caption{}
\label{fig:grad3}
\end{figure}

To more accurately determine the value of $\sigma$, we show in Figure \ref{fig:grad3}(a) a log-log plot of $(1-r_n/\mu)$ against ${n}.$ This is seen to be linear, and the gradient, calculated from the last two ratios $r_{99}$ and $r_{100}$ is -0.532. If one accepts that $\sigma$ is likely to be a simple rational number, the value $1/2$ is inescapable\footnote{With 200 terms the gradient estimate is improved to -0.522}.

Again, If we didn't know the value of $\mu,$ we could estimate $\sigma$ from the gradient of log-log plots of $r_{\sigma_n},$ see eqn. (\ref{eq:sig1}), or $a_{\sigma_n},$ see eqn. (\ref{eq:sig2}). It turns out, in contrast to the situation with the previous example, that estimators from (\ref{eq:sig2}) require hundreds of terms before a clear approach to the limit can be seen. However estimators from (\ref{eq:sig1}) are quite informative, though we still require 200 terms to draw any convincing conclusions. We show in Figure \ref{fig:grad3}(b) the estimate of $\sigma$ given by the gradient of the line joining the points $r_{\sigma_k}$ and $r_{\sigma_{k-1}},$ as $k$ ranges from 100 to 200. There is some curvature and oscillation in the plot, but clearly a limit of $1/2$ is attainable. 

Once again we see that if one wants to identify $\sigma$ as a simple fraction, likely to be $1/2$ or $1/3,$ then we have good evidence that it is $1/2.$ This can then be used in subsequent analysis.

In this plot, we used the fact that we knew $\mu=5$ exactly. If we didn't, we could linearly extrapolate the ratio plot in Figure \ref{fig:rat3}(b) which can be seen to be, plausibly, going to an ordinate value around 5. As in the preceding examples, we can also extrapolate the ratios using the Bulirsch-Stoer algorithm, this time with parameter $w=1/2.$ The results are shown in Table \ref{tab:BS3}. Only the first row behaves smoothly -- by which we mean monotonically. The monotonicity breaks down already in the second row. This is not totally surprising, as this series behaves slightly erratically, like the number of partitions of the integers. The number of interactions is not a fixed fraction of the length, and so low-order ratio plots are a little erratic going from one term to the next, though the global trend is uniform. One might estimate from this table that the limiting value was around $5.00$, and a brave person might conjecture that the limit is exactly 5. If one uses 200 terms instead of 100, the last entries are around 4.9995. 

\begin{table}
\caption{\label{tab:BS3}
Last seven entries in each row of the table of Bulirsch-Stoer extrapolants. Each successive row is the result of a successively higher degree of extrapolation. The available number of coefficients for extrapolation is $N$ (100 in this example). The highest order estimates, and presumably most precise, are in the last column.}
\begin{center}
\begin{tabular}{llllllll} \hline \hline
 L   &  T(L,N-L-6) & T(L,N-L-5) & T(L,N-L-4)& T(L,N-L-3) & T(L,N-L-2) & T(L,N-L-1) & T(L,N-L) \\
 
\hline

\hline
1 &5.06838354 &5.06758779 &5.06712079 &5.06606462& 5.06548413 &5.06480880 &5.06412456\\
2 &5.00326229 &5.00301024 &5.02650700 &4.98291297 &5.01522399 &5.00691875 &5.00510306 \\
\hline \hline
\end{tabular}
\end{center}
\end{table}

Assuming then that $\sigma=1/2,$ and $\mu=5,$ from (\ref{eq:half}) it follows that $$r_n/5 = 1 + \frac{ \log \mu_1}{2\sqrt{n}} + \frac{g+\frac{1}{8}\log^2 \mu_1}{n} + {\rm O}(n^{-3/2}). $$

As in example 1, in order to estimate $\mu_1$ and $g,$ we solve, sequentially, the pair of equations 
\BE \label{eq:fit2a}
r_j/5 = 1 + \frac{ c_1}{\sqrt{j}} + \frac{c_2}{j}, 
\EE
for $ j=k-1$ and $j=k,$ with $k$ ranging from 2 up to 200.

\begin{figure}[t!]
\centering
\subfigure[Estimates of parameter $c_1$ of (\ref{eq:fit2a}) against $\frac{1}{n}$.]{\includegraphics[width=7.3cm]{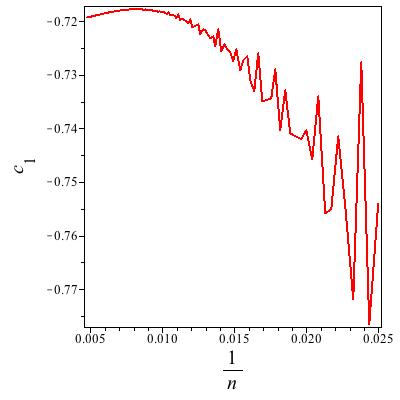}} 
\subfigure[Estimates of parameter $c_2$ of (\ref{eq:fit2a}) against $\frac{1}{\sqrt{n}}$.]{\includegraphics[width=7.3cm]{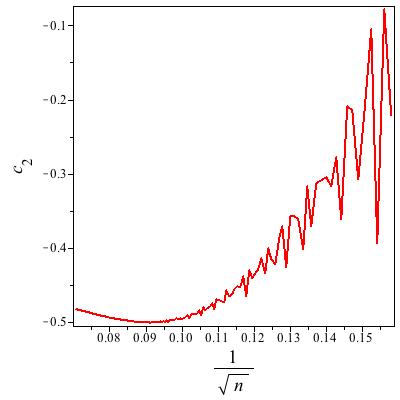}} 
\caption{}
\label{fig:fit3}
\end{figure}

The results are shown in figures \ref{fig:fit3}(a) and \ref{fig:fit3}(b). Simple visual extrapolation using the data points up to $n=100$ gives the estimate $c_1\approx -0.71.$ The plot for $c_2$ exhibits some curvature, and the best we can estimate is  $c_2 \approx -0.65$ from just 100 terms (note the negative gradient when $1/\sqrt{n} > 0.1$). From (\ref{eq:half}) these estimates imply $\log{\mu_1} \approx -1.42,$ and $g \approx -0.9.$ If we take 200 terms in the expansion instead of  100, we see that the plots have turning points at around $n=100,$ and that with $n=200$ our estimate of $c_2$ is close to $-0.4$. The 200 term series lets us make the more precise estimates $\log {\mu_1} \approx -1.44$ and $g\approx-0.66.$ It turns out that we need some 500 terms in the series before we can confidently estimate $g \approx -0.750$. We also estimated $\log{\mu_1} \approx -1.4396$ from a 500 terms series. This agrees with the analysis in \cite{OPB93} based on a 6000 term series, though they claim a more accurate estimate of $\mu_1,$ which is not given.

For this example the direct fitting method was somewhat less successful, as there is a substantial degree of oscillation in the plots of the various parameters, due to parity effects, as discussed above. Nevertheless, the results were useful, and if we use more than 100 terms, quite good accuracy can be achieved. But just using 100 terms, 
assuming $\sigma = 1/2$ and directly fitting to the remaining parameters, as described in subsection \ref{direct} above, we estimate
$c_1 \approx 1.61,$ implying $\mu \approx 5.003$ rather than the exact value of 5, $c_2 \approx -1.5$ so $\log(\mu_1) \approx -1.5$ rather than the more precise value $-1.439$, $c_3$ is in the range $[-1.4,-0.5],$ compared to the exact value of $-3/4,$ and $c_4 \approx 4$ rather than the actual value of around $0.8$.

If we assume that $\mu = 5$ and fit to the remaining three parameters, we estimate $c_2 \approx -1.415$ so $\log(\mu_1) \approx -1.415$ rather than the more precise value $-1.439$, $c_3 \approx -0.6,$ compared to the exact value of $-3/4,$ while $c_4 \approx 1.4$ is still a rather poor estimate. If we use 200 terms, the estimates improve to $c_2 \approx -1.44,$ $c_3 \approx -0.75,$ and $c_4 \approx 0.9$.

For IPDSAWs, as discussed above, the series do not behave smoothly at low order, due to the rather granular way the number of interactions increases with the length of the walk. So even the transformed series are not well-suited to analysis by the method of differential approximants. Nevertheless, the results of this approach are not without value. The critical point is estimated to be at $x_c \approx 0.2016,$ but with a second singularity very close by at $x \approx 0.208.$ The two singularities have associated exponents of opposite sign and varying magnitude, so that $g$ cannot be estimated this way.

We extrapolated the ratios of the coefficients of the transformed series using the Bulirsch-Stoer algorithm, with parameter $w=1.$ For IPDSAW the Bulirsch-Stoer extrapolants are, as expected, not monotonic, but do quickly settle down to values in the range $[4.9996,5.0010,$ in reasonable agreement with the exact value of $5.0$

In order to estimate the critical exponent $g,$ we extrapolated estimators $g_n = n^2\left ( 1-\frac{r_n}{r_{n-1}} \right )$ of the exponent against $1/n,$ as described at eqn. (\ref{eq:exp}). The extrapolants are not monotonic, but do quickly settle down to values in the range $[-1,-0.7].$ Using 250 terms in the series allows the much more precise estimate $g \approx -0.78,$ and with 500 terms that improves to $g = -0.75.$ The exact value is $g=-3/4.$ Note that these values were obtained without recourse to knowledge of the value of the growth constant $\mu.$ 

As with the previous two examples, we estimated the amplitude by dividing the coefficients by the terms we've identified in the asymptotic form.  The uncertainty in the value of $\mu_1$ again gives rise to a rather large uncertainty in the estimate of the amplitude $B.$ With the correct value of $\mu_1$ (or, rather, correct to four significant digits), this procedure gave $B \approx 2.22$ with a 100 term series. The estimate was slightly improved to $B \approx 2.210$ with a 700 term series.

This example represents the most difficult problem of this class, one in which the coefficients do not vary smoothly, and yet one has a singularity of the non-algebraic type that we are studying here. Despite this, a clear indication of the nature of the singularity was obtained, and reasonably accurate estimates of the critical parameters were also obtained, provided one has sufficient series coefficients.

\section{General methods to analyse such series.}
On the basis of these three examples, and others we have studied but not discussed at length here, we are now in a position to propose a method for analysing problems that may have coefficients of the assumed asymptotic form (\ref{asymp2}).

\begin{itemize}
\item Make a plot of the ratios against $1/n.$ If this plot is linear, or approaching linearity as $n$ increases, this is suggestive of an algebraic singularity.

\item Analyse the series by the method of differential approximants. If one obtains well-converged estimates of the position of the critical point(s) and exponent(s), and these are consistent with the ratio analysis, this is further evidence for an algebraic singularity. In those cases when the convergence is rapid and precise, as in the example of hexagonal SAPs in Section \ref{ana:hcsap}, one can have abundant confidence in this conclusion.

\item
If however the differential approximants are not well converged, and the associated exponent is poorly estimated and considered unlikely for a problem of the class being studied, then there is good reason to doubt that the underlying singularity is algebraic.

\item
If the ratio plot can be made linear by plotting the ratios against $1/\sqrt{n}$ or $1/n^{2/3},$ or other simple rational exponent $1/n^{1-\sigma},$ this is then  further evidence suggesting that the asymptotic form is not algebraic, but rather of the form (\ref{can1}). If one knows, or can accurately estimate, the radius of convergence $x_c,$ then denoting the ratio of successive coefficients by $r_n,$ a  plot of $\log {|r_n x_c-1|}$ against $\log{n}$ should give an estimate of the exponent needed to linearise the ratio plot. This should be consistent with that found by  choosing an exponent by trial and error to linearise the ratio plot. Otherwise, if $\mu$ is unknown, $\sigma$ can be estimated from the gradient of log-log plots of $r_{\sigma_n}$ and/or $a_{\sigma_n}$ against $n,$ where these quantites are defined in  eqns. (\ref{eq:sig1}) and (\ref{eq:sig2}) respectively.

\item 
The Bulirsch-Stoer or other appropriate extrapolation algorithm should be used to extrapolate the ratios of the coefficients in order to get a more reliable estimate of the growth constant $\mu.$ If the Bulirsch-Stoer algorithm is used, the parameter $w=1-\sigma,$ where $\sigma$ is estimated from the value needed to linearise the ratio plots. The estimate of $\mu$ can then be used in a log-log plot to refine the estimate of $\sigma.$ In this way a consistent pair of estimates of both $\mu$ and $\sigma$ can be obtained.

\item
Once the exponent $\sigma$ is well established, a four parameter fit to the assumed asymptotic form, as described in Section \ref{direct} should be conducted. Using the best estimate of the critical point, $1/\mu,$ this should be incorporated, and a three parameter fit tried. Naturally, parameter estimates by different methods should be mutually consistent.

\item 
Then an analysis which involves fitting successive pairs of ratios to $r_n \cdot x_c = 1 + \frac {c_1}{n^{1-\sigma} }+ \frac{c_2}{n}$ and extrapolating estimates of $c_1$ and $c_2$ should be performed. The estimates of $c_1$ and $c_2$ provide estimators of $\mu_1$ and $g.$ If one has very many terms, this can be taken further, and successive triples of ratios can be used to estimate the coefficients $c_i, \,\, i=1,2,3$ in 
$r_n \cdot x_c = 1 + \frac {c_1}{n^{1-\sigma} }+ \frac{c_2}{n}+\frac{c_3}{n^{2-2\sigma}}.$

\item
Finally, the series should be transformed to remove the factor $\mu_1^{n^\sigma},$ as discussed in Section \ref{sec:transform}, and the transformed series subject to standard analysis, such as ratio analysis, differential approximant analysis and ratio extrapolation by the Bulirsch-Stoer algorithm. The estimates of the critical parameters should be consistent with, and hopefully more precise than, those obtained by the other methods described.

\end{itemize}

\section{Conclusion}
We have described a number of methods to distinguish between coefficients of a generating function with algebraic singularities and those with more complicated non-algebraic singularities of the form $B \cdot \mu^n \cdot \mu_1^{n^\sigma} \cdot n^g.$ We have developed methods to identify coefficients in this latter class, and then shown  how to extract estimates for the critical parameters $B, \,\, \mu, \,\, \mu_1, \,\, \sigma, \,\, {\rm and} \,\, g.$ 

Our methods are based on extending the existing traditional methods for analysing algebraic singularities, the ratio method and the method of differential approximants. In the latter case we don't extend the method so much as its application. The fairly natural idea of directly fitting to the asymptotic form is also investigated, and found to be useful.

We illustrate these methods by applying them to three examples. They are the generating functions of fragmented permutations,  of compressed Dyck paths, and of IPDSAWs. In subsequent papers, of which \cite{CG14} is the first, we apply these ideas to previously unsolved problems.

\section{Acknowledgements}
I would like to thank Nick Beaton, Andrew Conway, Iwan Jensen, Einar Steingr\'imsson and Stu Whittington for careful reading of the manuscript, which resulted in very substantial improvement. I have benefited from discussions with Mireille Bousquet-M\'elou and Bruno Salvy on questions of singularities and asymptotic behaviour of coefficients, for which I am grateful. I'm also grateful to Robin Pemantle and Brendan McKay who sorted out the asymptotic behaviour of coefficients in the generating function for compressed Dyck paths, as discussed in Section 7. I would also like to thank the Australian Research Council who have supported this work through grant DP120100939.

\end{document}